\documentclass[aps,prd,superscriptaddress,notitlepage,showkeys,showpacs,10pt]{revtex4-1}

\usepackage{graphicx}
\usepackage{amsmath}
\usepackage{amssymb}
\usepackage{color}
\usepackage{bm}
\usepackage{hyperref}

\definecolor{purple}{rgb}{0.8,0,0.6}

\begin{document}
\title{Anomalous Maxwell equations for inhomogeneous chiral plasma}

\author{E. V.~Gorbar}
\affiliation{Department of Physics, Taras Shevchenko National Kiev University, Kiev, 03022, Ukraine}
\affiliation{Bogolyubov Institute for Theoretical Physics, Kiev, 03680, Ukraine}

\author{I. A.~Shovkovy}
\affiliation{College of Letters and Sciences, Arizona State University, Mesa, Arizona 85212, USA}

\author{S.~Vilchinskii}
\affiliation{Department of Physics, Taras Shevchenko National Kiev University, Kiev, 03022, Ukraine}
\affiliation{Departement de Physique Theorique Center for Astroparticle Physics, Universite de Geneve,
Quai E. Ansermet 24, 1211 Geneve 4, Switzerland}

\author{I.~Rudenok}
\affiliation{Department of Physics, Taras Shevchenko National Kiev University, Kiev, 03022, Ukraine}

\author{A.~Boyarsky}
\affiliation{Instituut-Lorentz for Theoretical Physics, Universiteit Leiden, Niels Bohrweg 2, 2333 CA Leiden, The Netherlands}

\author{O.~Ruchayskiy}
\affiliation{Discovery Center, Niels Bohr Institute, Blegdamsvej 17, DK-2100 Copenhagen, Denmark}

\begin{abstract}
Using the chiral kinetic theory we derive the electric and chiral current
densities in inhomogeneous relativistic plasma. We also derive
equations for the electric and chiral chemical potentials that close
the Maxwell equations in such a plasma. The analysis is done in the regimes
with and without a drift of the plasma as a whole. In addition
to the currents present in the homogeneous plasma (Hall current, chiral magnetic,
chiral separation, and chiral electric separation effects, as well as Ohm's current)
we derive several new terms associated with inhomogeneities of the
plasma. Apart from various diffusionlike terms, we find also new
dissipationless terms that are independent of relaxation time. Their origin
can be traced to the Berry curvature modifications of the kinetic theory.
\end{abstract}




\maketitle

\section{Introduction}

Nowadays there is a widespread interest in relativistic plasmas with a chiral asymmetry. This interest
is driven by a recent progress in the understanding of basic properties of numerous relativistic or pseudorelativistic
systems, ranging from Dirac and Weyl semimetals in condensed matter \cite{Vafek:2013mpa}, whose low-energy
quasiparticle excitations are described by the Dirac and Weyl equations, to strongly coupled quark-gluon
plasma created experimentally in heavy-ion collisions \cite{Liao:2014ava,Kharzeev:2015kna,Huang:2015oca},
and to the primordial plasma in the early Universe \cite{Rubinstein,Giovannini-review}. One of the 
most unusual features of a chiral relativistic plasma, which is absent in a conventional nonrelativistic 
one, is the possibility of a macroscopic realization of the celebrated quantum anomalies \cite{ABJ}. 
For example, in a chirally asymmetric plasma in a magnetic field $\mathbf{B}$ with an imbalance 
between the number densities of right- and left-handed fermions (described semirigorously by the 
chiral chemical potential $\mu_5$), the chiral anomaly induces a new type contribution to the 
electric current: $e\mathbf{j}=e^2\mu_5\mathbf{B}/(2\pi^2c)$~\cite{Vilenkin:80a,Redlich:1984md,
Tsokos:1985ny,Alekseev:1998ds}. The latter is known as the \emph{chiral magnetic effect} 
\cite{Fukushima:08,Kharzeev:2009fn}. The Maxwell equations, amended by such a contribution 
to the electric current, become \emph{anomalous Maxwell equations} \cite{Joyce:1997uy,
Giovannini:1997eg,Frohlich:2000en,Frohlich:2002fg,Boyarsky:2011uy,Tashiro:2012mf,
Giovannini:2013oga,Manuel:2015zpa,Boyarsky:2015faa,Hirono:2015rla}.

The inclusion of the anomalous currents drastically changes the self-consistent evolution of chiral charge 
densities and helical magnetic fields~\cite{Joyce:1997uy,Boyarsky:2011uy,Tashiro:2012mf,Manuel:2015zpa,
Boyarsky:2015faa,Hirono:2015rla}. Moreover, the nonlinear interactions due to anomalous processes
induce an effective mechanism for transferring the energy from magnetic modes with short wavelengths
(strongly affected by dissipation) to modes with longer wavelengths (and longer lifetimes) --- a phenomenon
similar to the inverse cascade in ordinary magnetohydrodynamics \cite{Christensson:2002xu}, but driven not by
turbulence. Depending on the chosen initial conditions, it was found that the helicity can be transferred from the fermions
to the magnetic fields or vice versa. These results support the suggestion made in Ref.~\cite{Avdoshkin:2014gpa}
that all degrees of freedom with nonvanishing axial charge are equally excited in the equilibrium.

The chiral anomaly in a relativistic matter exhibits itself not only via the chiral magnetic effect.
Even if a chiral asymmetry is absent, a nonzero magnetic field can induce an axial current
$\mathbf{j}_5=e\mu\mathbf{B}/(2\pi^2c)$ in a plasma with a nonzero electric chemical potential.
This phenomenon is known as the \emph{chiral separation effect} \cite{Metlitski:2005pr}. Another prediction
of the anomalous Maxwell equations in a relativistic plasma, that utilizes an interplay of the chiral
separation and chiral magnetic effects, is a new type of collective excitation known as
the \emph{chiral magnetic wave} \cite{Kharzeev:2010gd}. Further development of these ideas is
anomalous hydrodynamics, which contains new terms due to the chiral anomaly
\cite{Son:2009tf,Kharzeev:2011vv,Kalaydzhyan:11,Dubovsky:2011sj,Amado:2011zx,
Sadofyev:2010pr,Neiman:10,Megias:2013joa,Boyarsky:2015faa}.

The present study investigates the role of inhomogeneities in the chiral plasma evolution.
Since the chiral anomaly relation is local, it is clear that the electric and chiral chemical potentials
should be inhomogeneous just like the helical electromagnetic fields. An important question is whether
additional contributions to the electric current exist in the inhomogeneous case. The authors of
Ref.~\cite{Giovannini:1997eg} postulated the absence of such currents, while a different set of
equations was proposed in a recent study \cite{Boyarsky:2015faa}. The present paper derives
such inhomogeneous currents in a systematic way. Our starting point in the analysis is the chiral
kinetic theory \cite{Son:2012zy,Stephanov:2012ki,Chen:2014cla,Manuel:2014dza}.

This paper is organized as follows. We briefly review the chiral kinetic theory and kinetic equations in
Sec.~\ref{CKT}. The expressions for the electric and axial currents and equations for the local
equilibrium chemical potentials in inhomogeneous chiral plasma in the drifting state are derived
in Sec.~\ref{equation-mu5}. The electric and chiral currents to the second order in electromagnetic
field and derivatives are calculated in Sec.~\ref{perturbation-fields} in the case where neutral particles
exert a substantial drag on the system. The summary and conclusions are given in
Sec.~\ref{Conclusion}. Some table integrals and useful relations are collected in the Appendix.

\section{Chiral kinetic theory}
\label{CKT}

The chiral kinetic theory describes a time evolution of the one-particle distribution functions
$f_{\lambda}(t,\mathbf{x},\mathbf{p})$ for the right- ($\lambda=+$) and left-handed ($\lambda=-$)
fermions. It was recently suggested \cite{Xiao,Duval:2005vn,Son:2012zy} that chiral fermions in external electromagnetic fields are described by the \emph{chiral kinetic theory} given by \cite{Son:2012zy,Stephanov:2012ki,Chen:2014cla,Manuel:2014dza}
\begin{equation}
\frac{\partial f_{\lambda}}{\partial t }+\frac{1}{1+\frac{e}{c}\mathbf{B}\cdot\mathbf{\Omega}_{\lambda}}
\left[\Big(e\mathbf{E}+\frac{e}{c}\mathbf{v}\times \mathbf{B}
+\frac{e^2}{c}(\mathbf{E}\cdot\mathbf{B})\mathbf{\Omega}_{\lambda}\Big)\cdot\frac{\partial f_{\lambda}}{\partial \mathbf{p}}
+\Big(\mathbf{v}+e\mathbf{E}\times\mathbf{\Omega}_{\lambda}
+\frac{e}{c}(\mathbf{v}\cdot\mathbf{\Omega}_{\lambda})\mathbf{B}\Big)
\cdot\frac{\partial f_{\lambda}}{\partial \mathbf{x}}\right]
=I_{\rm coll},
\label{kinetic-equation}
\end{equation}
where the factor $(1+e\mathbf{B}\cdot\mathbf{\Omega}_{\lambda}/c)^{-1}$ accounts for the
correct phase-space volume \cite{Xiao,Duval:2005vn} and $\mathbf{\Omega}_{\lambda}
=\lambda\mathbf{p}/(2|\mathbf{p}|^3)$ is the \emph{Berry curvature} \cite{Berry}. The Berry
curvature is a crucial ingredient in the chiral kinetic theory that allows one to capture
the fermionic nature of particles \cite{Son:2012zy,Stephanov:2012ki,Chen:2014cla,Manuel:2014dza}. The 
group velocity is defined from the quasiparticle energy as follows: $\mathbf{v}\equiv 
\partial \epsilon_{\mathbf{p}}/\partial \mathbf{p}$. By imposing the constraint of the 
Lorentz invariance in Ref.~\cite{Son:2012zy}, it was suggested that the dispersion relation 
for chiral fermions in a magnetic field $\mathbf{B}$ should be taken in the form 
$\epsilon_{\mathbf{p}} =c|\mathbf{p}|-\lambda e\mathbf{p}\cdot\mathbf{B}/|\mathbf{p}|^2$,
which is valid to linear order in the field when $|e\mathbf{B}|/(c\mathbf{p}^2) \ll 1$. Interestingly, 
however, such a definition for $\epsilon_{\mathbf{p}}$ may be problematic because the absolute value of 
the resulting group velocity, $v=c \sqrt{1+2e \left(\mathbf{B} \cdot \mathbf{\Omega}_{\lambda}\right)/c 
+O(\mathbf{B}^2)}$, appears to be larger than the speed of light when 
$(\mathbf{B} \cdot \mathbf{\Omega}_{\lambda})>0$. In the present study, we will use the dispersion 
relation $\epsilon_{\mathbf{p}}=c|\mathbf{p}|$ for which the group velocity equals $c$.

Equation (\ref{kinetic-equation}) is the kinetic equation for the distribution function
$f^i_{\lambda}$ of particles ($i=p$). A separate equation can be written down for antiparticles
($i=a$). The corresponding equation can be obtained by simply replacing $e \to -e$
and $\lambda \to -\lambda$ in Eq.~(\ref{kinetic-equation}). Below, when there is no risk
of confusion, we will omit index $i$ and assume that the expressions are given for particles.
In the end, the results for charge densities and current densities will have to contain both
particle and antiparticle contributions.

The term on the right-hand side of the kinetic equation (\ref{kinetic-equation}) is a collision integral.
In the simplest approximation, one can take $I_{\rm coll}=0$. This corresponds to the so-called
collisionless limit, which is useful when the collective particle dynamics is driven primarily by
averaged electromagnetic fields. One of the simplest approximations beyond the collisionless 
limit is the relaxation-time approximation with 
$I_{\rm coll}=-(f_{\lambda}-f^{\rm (eq)}_{\lambda})/\tau$ \cite{Son-Spivak,Satow},
where $\tau\sim 1/\left[e^4T\ln(1/|e|)\right]$ \cite{Arnold} is the relaxation time and
$f^{\rm (eq)}_{\lambda}$ is the local equilibrium distribution function. In the absence 
of electromagnetic fields, it is the standard Fermi-Dirac distribution function
\begin{equation}
f^{\rm (eq)}_{\lambda}(t,\mathbf{x},\mathbf{p})
=\frac{1}{e^{[\epsilon_{\mathbf{p}}-\mu_{\lambda}(t,\mathbf{x})]/T}+1} ,
\label{equilibrium-function}
\end{equation}
where $\epsilon_\mathbf{p} = c|\mathbf{p}|$. The corresponding equilibrium distribution function 
for antiparticles is obtained by replacing $\mu_{\lambda} \to -\mu_{\lambda}$. Here we introduced
the notation for the chemical potentials of the right- and left-handed fermions,
$\mu_{\lambda}(t,\mathbf{x})=\mu(t,\mathbf{x})+\lambda\mu_5(t,\mathbf{x})$. It should be noted that,
in the case of local equilibrium, the temperature $T$ in distribution functions could also depend
on the space-time coordinates, $T(t,\mathbf{x})$. However, in order to simplify our analysis below,
we will neglect such a dependence in what follows. In the presence of electromagnetic fields,
the choice of $f^{\rm (eq)}_{\lambda}(t,\mathbf{x},\mathbf{p})$ is a delicate issue and we will discuss
it in more detail below.

In a general case, the local equilibrium chemical potentials $\mu_{\lambda}$
evolve with time. Therefore, one of the central and crucial
points of our analysis is the evolution equations for $\mu_{\lambda}$ which we derive
from the kinetic equation. Integrating the left-hand side of the kinetic equation
(\ref{kinetic-equation}) over momentum leads to the continuity equations for the
electric and chiral currents, where the latter equation includes the chiral anomaly term.
This means that in order that the particles densities be conserved it is necessary that the
integral of the collision integral over momentum be equal to zero. This requirement 
will play a crucial role in our analysis below.

A collision integral $I^{\rm BGK}_{\rm coll}=-(f_{\lambda}-\frac{n_{\lambda}}{n_{\lambda}^{(0)}}f_{\lambda}^{(0)})/\tau$
of the Bhatnagar-Gross-Krook (BGK)-type \cite{BGK} was used in Ref.~\cite{Manuel:2015zpa}. Here $n_{\lambda}$ is
a local fermion number density, $f^{(0)}_{\lambda}$ is a given distribution function (for example, in the
analysis of nonrelativistic particle dynamics in Ref.~\cite{BGK}, a Maxwell velocity distribution function
was used), and $n_\lambda^{(0)}$ is determined by $f^{(0)}_{\lambda}$. Clearly, such a collision
integral automatically conserves the particle number and agrees with the chiral anomaly relation.
We found that when studying the evolution of  electromagnetic fields and the chiral asymmetry in
the presence of strong magnetic fields, where the chiral chemical potential evolves with time, it is
crucial to use the local equilibrium function $f^{\rm (eq)}_{\lambda}$ rather than a fixed $f^{(0)}_{\lambda}$,
otherwise, the kinetic equation is not entirely consistent. We checked, however, that the results obtained
for the BGK-type collision integral with $f^{\rm (eq)}_{\lambda}$ instead $f^{(0)}_{\lambda}$ are not much
different from those found in the relaxation-time approximation described above. Since the relaxation-time
approximation with $I_{\rm coll}=-(f_{\lambda}-f^{\rm (eq)}_{\lambda})/\tau$ is slightly simpler, we will use it in our
analysis below.

The evolution of electric and magnetic fields is determined by the Maxwell equations
\begin{eqnarray}
\bm{\nabla}\cdot\mathbf{E} &=& 4\pi e n,
\label{Poisson-equation} \\
\bm{\nabla}\times\mathbf{E} &=& -\frac{1}{c}\frac{\partial \mathbf{B}}{\partial t},
\label{vector-equation}\\
\bm{\nabla}\cdot\mathbf{B} &=& 0,
\label{scalar-equation}\\
\bm{\nabla}\times\mathbf{B} &=& \frac{4\pi}{c}e\mathbf{j}+\frac{1}{c}\frac{\partial \mathbf{E}}{\partial t}.
\label{Maxwell-equations}
\end{eqnarray}
By definition, the electric charge density is given by $e n(\mathbf{x})=\sum_{i} \sum_{\lambda=\pm}
e_i n^{i}_{\lambda}(\mathbf{x})$, where the sum over $i$ includes the contributions due to particles
($i=p$) and antiparticles ($p=a$). Note that the latter comes with the opposite sign. The
number density of particles of a given chirality $\lambda$ is given by
\begin{equation}
n_{\lambda}(\mathbf{x})=\int\frac{d^3p}{(2\pi)^3}
\left(1+\frac{e}{c}\mathbf{B}\cdot\mathbf{\Omega}_{\lambda}\right)f_{\lambda}(\mathbf{p},\mathbf{x}).
\label{charge-density}
\end{equation}
Here, the factor $1+e\mathbf{B}\cdot\mathbf{\Omega}_{\lambda}/c$ in the integrand takes care of the correct
phase-space volume. After multiplying the kinetic equation (\ref{kinetic-equation})
by $1+e\mathbf{B}\cdot\mathbf{\Omega}_{\lambda}/c$, integrating over momentum $\mathbf{p}$, and using
the Maxwell equations, we obtain the following relation:
\begin{equation}
\partial_t n_{\lambda}+\bm{\nabla}\cdot\mathbf{j}_{\lambda}=-\frac{e^2}{c}\int\frac{d^3p}{(2\pi)^3}
\Big(\mathbf{\Omega}_{\lambda}\cdot\nabla_{\mathbf{p}}f_{\lambda}\Big)\mathbf{E}\cdot\mathbf{B}
=\frac{\lambda  e^2\mathbf{E}\cdot\mathbf{B}f_{\lambda}(\mathbf{p}=0)}{4\pi^2c},
\label{continuity-equation}
\end{equation}
where we integrated by parts in the last equality and used the following identity for the Berry curvature
$\nabla_{\mathbf{p}}\cdot\mathbf{\Omega}_{\lambda}=2\pi\lambda \delta^3(\mathbf{p})$. The electric
current density is given by $e \mathbf{j}(\mathbf{x})=\sum_{i} \sum_{\lambda=\pm}
e_i \mathbf{j}^{i}_{\lambda}(\mathbf{x})$, where the contribution due to particles of a given chirality is
determined by \cite{Son-Spivak,Son:2012zy,Stephanov:2012ki}
\begin{equation}
\mathbf{j}_{\lambda}= \int\frac{d^3p}{(2\pi)^3}\,\left(\mathbf{v}+e\mathbf{E}\times\mathbf{\Omega}_{\lambda}
+\frac{e}{c}\mathbf{B}(\mathbf{v}\cdot\mathbf{\Omega}_{\lambda})\right)\,f_{\lambda}
+ \mathbf{j}^{\rm (curl)}_{\lambda}.
\label{electric-current}
\end{equation}
We should note that the definition of the current in Ref.~\cite{Son-Spivak} differs from
that in Ref.~\cite{Son:2012zy}. The latter has an extra term $\mathbf{j}^{\rm (curl)}_{\lambda}$, the
explicit form of which can be obtained by integrating the definition of current in Ref.~\cite{Son:2012zy}
by parts,
\begin{equation}
\mathbf{j}^{\rm (curl)}_{\lambda}= \bm{\nabla}  \times \int\frac{d^3p}{(2\pi)^3}
f_{\lambda}\epsilon_{\mathbf{p}}\mathbf{\Omega}_{\lambda} .
\label{mag-current}
\end{equation}
As we see, this contribution is a total curl and, thus, does not contribute to the continuity equation.
However, such a current affects the Maxwell equation (\ref{Maxwell-equations}). From the structure
of the corresponding equation and the structure of the current, we see that the integral on the right
hand on Eq.~(\ref{mag-current}) plays the role of a magnetization. Then, the current itself can be
viewed as a ``magnetization" current \cite{footnote}.

It is instructive to emphasize that, in the case of hot relativistic plasmas, it is essential that the
complete expressions for the electric/chiral charge densities, as well as
the corresponding currents contain the contributions of both particles and antiparticles. This is
in contrast to the case of dense relativistic plasmas at low temperatures, $T\ll |\mu_\lambda|$,
when the contributions of antiparticles are exponentially suppressed and, therefore, could be safely neglected.
In the high-temperature regime, $T\gg |\mu_\lambda|$, which is of prime
importance in cosmology, the antiparticle number densities are given by the same expressions as
in Eq.~(\ref{charge-density}), but in terms of the antiparticle distribution functions $f^{\rm a}_{\lambda}$.
Taking into account that antiparticles carry the opposite electric charge, we will find that, as
expected, the corresponding high-temperature plasma will be almost neutral. Perhaps even more importantly,
antiparticles will contribute approximately as much as particles to the electric current and, thus, will
effectively double the result.

It is interesting to note that taking the contribution of antiparticles into account is critical also in order to obtain
the correct chiral anomaly relation from Eq.~(\ref{continuity-equation}). The right-hand side of the
corresponding equation for particles is proportional to the local equilibrium distribution function
$f^{p(\rm{eq})}_{\lambda}(\mathbf{p}=0)$, which depends on $\mu_{\lambda}$ and temperature.
This seems to be at odds with the topological nature of the corresponding relation. The same
is true for antiparticles. However, in view of the identity $\sum_i f_{\lambda}^{i}(\mathbf{p}=0)=1$,
the total chiral current $\mathbf{j}_5=\sum_i\sum_{\lambda=\pm}\lambda\,\mathbf{j}^i_{\lambda}$
does satisfy the conventional continuity equation with the correct anomalous term
$e^2\mathbf{E}\cdot\mathbf{B}/(2\pi^2c)$. Note that, in the low-temperature regime, the
correct result is saturated almost exclusively by the contribution of particles. (Of course,
it is easy to check that the electric current $e\mathbf{j}=\sum_i\sum_{\lambda=\pm}
e_i\mathbf{j}^i_{\lambda}$ satisfies the usual nonanomalous continuity equation.)

The complete set of the Maxwell equations (\ref{Poisson-equation}) through (\ref{Maxwell-equations}),
the kinetic equation (\ref{kinetic-equation}) in the relaxation-time approximation supplemented by the
definitions of the number densities (\ref{charge-density}) and currents (\ref{electric-current}) form
a system of self-consistent equations for the one-particle distribution functions of the left- and
right-handed fermions and electromagnetic fields. Therefore, to study, for example, the evolution
of inhomogeneous magnetic fields and chiral asymmetry in magnetized plasma, one should solve
the corresponding system of equations. The corresponding task is formidable. In order to simplify it,
we will derive an approximate set of equations in the case where electromagnetic fields are weak.

Before proceeding to the derivation of a complete set of equations that describe the evolution
of an inhomogeneous chiral plasma coupled to electromagnetic fields, it is instructive to identify
generic classes of such plasmas, depending on their composition and underlying dynamics.
In particular, as will become clear below, an important role in the analysis is played by a
possible presence of additional neutral particles and their interactions with the charged carriers
of the plasma.

In order to understand the role of neutral particles better, let us remind a very special property of
a plasma that contains no such particles. When a configuration of perpendicular electric and
magnetic fields, $\mathbf{E}\perp \mathbf{B}$ (assuming only that $E<B$), is applied, such
a plasma drifts as a whole with the velocity perpendicular to both electric and magnetic fields 
\cite{Krall}
\begin{equation}
  \label{eq:1}
  \bar{\mathbf{v}}=c\frac{\mathbf{E}\times\mathbf{B}}{B^2},
\end{equation}
see also Sec.~\ref{equation-mu5} below.
It is crucial that the drift velocity $\bar{\mathbf{v}}$ does {\em not} depend on the specific values of charges
(and masses, if present) of particles. In fact, despite the drift, the plasma is in perfect equilibrium.
It is described by a boosted, rather than the usual form of the Fermi-Dirac distribution function.
(This can be also understood from a different angle: there is no electric field in the boosted
reference frame, moving with the velocity $\bar{\mathbf{v}}$ with respect to the laboratory frame.)

It should be clear that the above-mentioned drifting state of the plasma should be profoundly
affected whenever a background is present that exerts a drag on the system. In solid state materials,
for example, the corresponding background could be due to the lattice of ions or impurities. In other
plasmas, it could be due to neutral particles, which are not affected by electromagnetic fields directly.
In the latter case, of course, it is assumed that the time scale for the neutral component to develop its
own drift (via the interaction with the charged particles) is much longer than the characteristic time
scales for the evolution of electromagnetic fields and inhomogeneities.

In this paper, we will discuss both cases with and without a drift of the plasma as a whole.
Perhaps one of the best realistic examples of the plasma that is subject to the drift is a relativistic
QED electron-positron plasma. An example of a plasma where the drift may not fully develop to
involve the background of neutral particles is the quark-gluon plasma at sufficiently high temperatures.
The (electromagnetically) neutral particles in the latter case are gluons. In the case of chiral plasmas in
Dirac/Weyl semimetals, the background is due to the lattice ions or impurities that do not develop any drift at all.

\section{Chiral plasma in the drifting state: expansion in $\mathbf{E}_{\parallel}$,
$\mathbf{B}$, and derivatives}
\label{equation-mu5}

In this section, we will derive the equations describing the evolution of a weakly inhomogeneous chiral
plasma and electromagnetic fields without any additional components (e.g., neutral particles, pinned impurities,
or ion lattices) present that could exert a substantial drag on the charged particles. If additional components
are present, it is assumed that their interaction with charged particles is negligible and has no qualitative
effect on the electromagnetic dynamics. We will use the kinetic equation (\ref{kinetic-equation}) in the relaxation
time approximation. The distribution functions for charged particles ($i=p$) and their antiparticles ($i=a$)
satisfy the following equations:
\begin{eqnarray}
&&\left(1+\frac{e_i}{c}\mathbf{B}\cdot\mathbf{\Omega}^i_{\lambda}\right)
\frac{\partial f^i_{\lambda}}{\partial t }+
\left(e_i\mathbf{E}+\frac{e_i}{c}\mathbf{v}\times \mathbf{B}
+\frac{e^2_i}{c}(\mathbf{E}\cdot\mathbf{B})\mathbf{\Omega}^i_{\lambda}\right)
\cdot\frac{\partial f^i_{\lambda}}{\partial \mathbf{p}}
\nonumber\\
&&+\left(\mathbf{v}+e_i\mathbf{E}\times\mathbf{\Omega}^i_{\lambda}
+\frac{e_i}{c}(\mathbf{v}\cdot\mathbf{\Omega}^i_{\lambda})\mathbf{B}\right)
\cdot\frac{\partial f^i_{\lambda}}{\partial \mathbf{x}}
=-\frac{1}{\tau} \left(1+\frac{e_i}{c}\mathbf{B}\cdot\mathbf{\Omega}^i_{\lambda}\right)\left( f^i_{\lambda}-f_{\lambda}^{i\,\rm (eq)}\right),
\label{approxBoltz-collision-integral}
\end{eqnarray}
where $e_i=e$, $\Omega^i_{\lambda}=\Omega_{\lambda}$ for particles and
$e_i=-e$, $\Omega^i_{\lambda}=-\Omega_{\lambda}$ for antiparticles.

In the case of a plasma made of only charged degrees of freedom, the standard Fermi-Dirac distribution
(\ref{equilibrium-function}) may not be the best choice for the zeroth order of the equilibrium distribution
function. Ideally, one would want to capture the drift of the plasma as a whole by a modified distribution
function. The line of arguments that allows one to obtain the corresponding function is well known.

Before considering a general configuration of electromagnetic fields, let us start by recalling that a field
configuration with constant $\mathbf{E}\perp \mathbf{B}$ (and $E<B$) has no dissipative effects on the
plasma. (Of course, this will change when the parallel component $\mathbf{E}_\parallel$ is added later 
as a perturbation.)
The whole system simply drifts with the velocity $\bar{\mathbf{v}}$ \cite{Krall}, see Eq.~\eqref{eq:1}. 
Of course, this is connected with the fact that, in the reference frame $K^{\prime}$, moving with the 
drift velocity $\bar{\mathbf{v}}$ with respect to the laboratory reference frame $K$, the electric field 
is absent \cite{field-theory}. In the absence of electric field in the $K^{\prime}$ frame, the equilibrium
state is naturally described by the standard Fermi-Dirac distribution function in the relativistic notation
\cite{Cerignani} (for the sake of simplicity, we suppress the particle/antiparticle index $i$)
\begin{equation}
f^{\prime\,\rm (eq)}_{\lambda}=\frac{1}{\exp\left(\frac{p^{\prime\nu} u^{\prime}_\nu -\mu^{\prime}_{\lambda}}{T^{\prime}} \right)+1}
=\frac{1}{\exp\left(\frac{\epsilon_{\mathbf{p}^{\prime}}-\mu^{\prime}_{\lambda}}{T^{\prime}} \right)+1},
\label{f-K-prime}
\end{equation}
where $p^{\prime\,\nu}=(\epsilon_{\mathbf{p}^{\prime}}/c,\mathbf{p}^{\prime})$ and $u^{\prime\,\nu}
\equiv (u^{\prime\,0}, \mathbf{u}^{\prime})=(c,0,0,0)$ is the proper velocity. We emphasize that this consideration 
assumes that either no neutral particles are present or that their interaction is too weak to change the
evolution of electromagnetic fields in the inhomogeneous plasma.

By performing the inverse Lorentz transformation in Eq.~(\ref{f-K-prime}), we easily find the equilibrium
distribution function in the laboratory frame, i.e.,
\begin{equation}
f^{\rm (\rm{eq})}_{\lambda}
=\frac{1}{\exp\left(\frac{\epsilon_{\mathbf{p}}-\mathbf{p}\cdot\bar{\mathbf{v}}-\mu_{\lambda} }{T} \right)+1},
\label{transformed-distribution-function}
\end{equation}
where  $u^{\nu} \equiv (u^0, \mathbf{u})=\left(c/\sqrt{1-(\bar{v}/c)^2},\bar{\mathbf{v}}/\sqrt{1-(\bar{v}/c)^2}\right)$,
$T=T^{\prime}\sqrt{1-(\bar{v}/c)^2}$, and $\mu_{\lambda}=\mu^{\prime}_{\lambda}\sqrt{1-(\bar{v}/c)^2}$. Note
that $T^{\prime}$ and $\mu^{\prime}_{\lambda}$ are Lorentz scalars that have the meaning of the temperature 
and chemical potentials in the local rest frame of the plasma. From the form of the distribution function 
(\ref{transformed-distribution-function}), the parameters $T$ and $\mu_{\lambda}$ appear to play the role of 
the temperature and chemical potentials in the laboratory reference frame. Such an interpretation of $T$ and $\mu_{\lambda}$
should be used with caution, however, because the plasma is not stationary with respect to the laboratory frame.
It is not difficult to check that the distribution function (\ref{transformed-distribution-function}) is indeed a stationary 
solution of the kinetic equation (\ref{approxBoltz-collision-integral}) for constant perpendicular electric and magnetic 
fields.

The velocity $\bar{\mathbf{v}}=c\mathbf{E}\times\mathbf{B}/B^2$ is known in the plasma physics as the drift
velocity \cite{Krall} because the motion of charged particles in constant perpendicular electric and magnetic
fields is the superposition of circular motion around a point called the guiding center and a drift of this point
with the velocity $\bar{\mathbf{v}}$. It is crucially important for us that the drift velocity does not depend on
the charges (and masses) of particles. In fact, this remains true also in a nonrelativistic plasma.

It should be emphasized that the plasma drift is well defined only when $E_\perp < B$. In this case
the drift speed $\bar{v}=c E_\perp /B$ is smaller than the speed of light $c$. In the opposite case, $E_\perp > B$,
there is no reference frame $K^\prime$, in which the perpendicular component of the electric field vanishes.

By using function (\ref{transformed-distribution-function}) as the zeroth order approximation for the distribution
function, let us proceed to the analysis of the general case when the parallel component of the electric field
$\mathbf{E}_\parallel$ is also present and drives the system out of equilibrium. We will seek the
solution for $f_{\lambda}$ in the form of an expansion, i.e.,
\begin{equation}
f_{\lambda}=f^{\rm (eq)}_{\lambda}+\delta f^{(1)}_{\lambda}+\cdots,
\label{ansatz-transformed}
\end{equation}
where $\delta f^{(1)}_{\lambda}$ defines a deviation from the local equilibrium to the first order in the parallel
electric field $\mathbf{E}_{\parallel}$, magnetic field $\mathbf{B}$, and the first derivatives of electromagnetic fields and chemical
potentials.

By substituting expansion (\ref{ansatz-transformed}) into the kinetic equation (\ref{approxBoltz-collision-integral})
and keeping only the terms up to linear order, we obtain
\begin{equation}
\frac{\bar{D}_{\lambda}}{T}\frac{\partial (\mu_{\lambda}+\mathbf{p}\cdot\bar{\mathbf{v}})}{\partial t }
-\frac{\bar{D}_{\lambda}}{T} e\frac{\left(\mathbf{E}\cdot \mathbf{B} \right)\left(\mathbf{v}\cdot \mathbf{B} \right)}{B^2}
+ \frac{\bar{D}_{\lambda}}{T} \mathbf{v}\cdot \frac{\partial (\mu_{\lambda}+\mathbf{p}\cdot\bar{\mathbf{v}})}{\partial \mathbf{x}}
+\frac{\delta f^{(1)}_{\lambda}}{\tau} =0,
\end{equation}
where we used the notation $\bar{D}_{\lambda}(\mu_{\lambda})
\equiv f^{\rm (eq)}_{\lambda}\left(1-f^{\rm (eq)}_{\lambda}\right)
=e^{(cp-\mathbf{p}\cdot\bar{\mathbf{v}}-\mu_{\lambda})/T}
/(e^{(cp-\mathbf{p}\cdot\bar{\mathbf{v}}-\mu_{\lambda})/T}+1)^2$. Note that the drift velocity $\bar{\mathbf{v}}$
is defined in terms of electromagnetic fields and, thus, may depend on the spacetime coordinates. By solving the above equation, we obtain
\begin{equation}
\delta f^{(1)}_{\lambda}=\frac{\tau \bar{D}_{\lambda}}{T} \left(
e\frac{\left(\mathbf{E}\cdot \mathbf{B} \right)\left(\mathbf{v}\cdot \mathbf{B} \right)}{B^2}
-\mathbf{v}\cdot \frac{\partial (\mu_{\lambda}+\mathbf{p}\cdot\bar{\mathbf{v}})}{\partial \mathbf{x}}
- \frac{\partial (\mu_{\lambda}+\mathbf{p}\cdot\bar{\mathbf{v}})}{\partial t } \right) .
\label{f1-new}
\end{equation}
By making use of this first-order correction to the distribution function $f^{i\,\rm{(eq)}}_{\lambda}$, we can
now calculate the charge densities and current densities to the same order.

The first-order result for the chiral charge densities reads
\begin{eqnarray}
n_\lambda  &=& \sum_{i} {\rm sign}(e_i)
\int \frac{d^3p}{(2\pi)^3}\left(f^{i\,\rm{(eq)}}_{\lambda}+ \delta f^{i, (1)}_{\lambda}
+\frac{e_i}{c} (\mathbf{B}\cdot\mathbf{\Omega}^i_{\lambda}) f^{i\,\rm{(eq)}}_{\lambda}
\right)
\nonumber\\
&=&n^{(0)}_\lambda
- \tau \frac{\partial n^{(0)}_\lambda }{\partial \mu_{\lambda}} \left(
\bar{\mathbf{v}}\cdot \frac{\partial \mu_{\lambda}}{\partial \mathbf{x}} + \frac{\partial \mu_{\lambda}}{\partial t } \right)
 -  \tau n^{(0)}_\lambda
\left[  (\bm{\nabla}\cdot\bar{\mathbf{v}})
+\frac{4 \bar{v} }{(c^2-\bar{v}^2)} \left( \bar{\mathbf{v}}\cdot\frac{\partial \bar{v}}{\partial \mathbf{x}}
+ \frac{\partial \bar{v}}{\partial t }\right)
\right],
\label{density-lambda-1}
\end{eqnarray}
where, by definition,
\begin{equation}
n^{(0)}_\lambda = \frac{\mu_{\lambda}^3+\pi^2 T^2\mu_{\lambda} }{6\pi^2 c^3\left[1-(\bar{v}/c)^2\right]^2}.
\label{density-lambda-00}
\end{equation}
In deriving expression (\ref{density-lambda-1}), we  took into account that $(\mathbf{B}\cdot \bar{\mathbf{v}}) = 0$.

To the same first order, the current densities are given by
\begin{eqnarray}
\mathbf{j}_{\lambda} &=&  \sum_{i} {\rm sign}(e_i)
\int\frac{d^3p}{(2\pi)^3}\,\left[ \left(\mathbf{v} +e_i\mathbf{E}\times\mathbf{\Omega}^i_{\lambda}
+\frac{e_i}{c}\mathbf{B}(\mathbf{v}\cdot\mathbf{\Omega}^i_{\lambda})\right)f^{i\,\rm{(eq)}}_{\lambda}
+ \bm{\nabla}  \times \left(
\epsilon_{\mathbf{p}}\mathbf{\Omega}_{\lambda} f^{i\,\rm{(eq)}}_{\lambda} \right) 
+ \mathbf{v}  \delta f^{i, (1)}_{\lambda}
 \right] \nonumber\\
&=& \mathbf{j}^{\rm (Hall)}_{\lambda} +\mathbf{j}^{(1)}_{\lambda}+\mathbf{j}^{(2)}_{\lambda}
+\mathbf{j}^{\rm (curl)}_{\lambda} +\mathbf{j}^{(3)}_{\lambda},
\label{current-lambda-1}
\end{eqnarray}    
where the first four contributions are independent of the relaxation time, while the last one,
$\mathbf{j}^{(3)}_{\lambda}$, is linear in $\tau$. The Hall current has the standard form, i.e.,
\begin{equation}
\mathbf{j}^{\rm (Hall)}_{\lambda} =
\sum_{i} {\rm sign}(e_i) \int\frac{d^3p}{(2\pi)^3}\,   \mathbf{v} f^{i\,\rm{(eq)}}_{\lambda}
= c n^{\rm{(0)}}_\lambda  \frac{\mathbf{E} \times \mathbf{B}}{B^2}.
\label{Hall-drift}
\end{equation}
The explicit expressions of the other two nondissipative terms are given by
\begin{eqnarray}
\mathbf{j}^{(1)}_{\lambda}&=& \sum_{i} {\rm sign}(e_i)
\int\frac{d^3p}{(2\pi)^3}\, e_i (\mathbf{E}\times\mathbf{\Omega}^i_{\lambda} ) f^{i\,\rm{(eq)}}_{\lambda}
=\frac{\lambda e \mu_{\lambda}}{4 \pi^2 \bar{v}^3} (\mathbf{E}\times\bar{\mathbf{v}})
\left(\frac{c}{2}\ln\frac{c+\bar{v}}{c-\bar{v}}-\bar{v}\right) , \\
\mathbf{j}^{(2)}_{\lambda}&=& \sum_{i} {\rm sign}(e_i)
\int\frac{d^3p}{(2\pi)^3}\, \frac{e_i}{c}\mathbf{B}(\mathbf{v}\cdot\mathbf{\Omega}^i_{\lambda}) f^{i\,\rm{(eq)}}_{\lambda}
= \frac{\lambda e \mu_{\lambda}}{8\pi^2 \bar{v}} \mathbf{B} \ln\frac{c+\bar{v}}{c-\bar{v}}.
\end{eqnarray}
By making use of the definition for the drift velocity, it is easy to check that, in addition to a perpendicular
component, the current $\mathbf{j}^{(1)}_{\lambda}$ also contains a contribution parallel to the
magnetic field. By combining the corresponding parallel component with the other current, $\mathbf{j}^{(2)}_{\lambda}$,
we obtain the usual currents of the chiral magnetic and chiral separation effects
\begin{equation}
\mathbf{j}^{\rm (CME)}_{\lambda} = \frac{\lambda e \mu_{\lambda}}{4 \pi^2 c} \mathbf{B}.
\label{CME-current}
\end{equation}
The remaining contribution is perpendicular to the magnetic field. Its explicit form reads
\begin{eqnarray}
\mathbf{j}^{(\perp)}_{\lambda}&=& \mathbf{j}^{(1)}_{\lambda} + \mathbf{j}^{(2)}_{\lambda}
-\mathbf{j}^{\rm (CME)}_{\lambda}
= \frac{\lambda e \mu_{\lambda}}{4\pi^2 c}  \mathbf{E}_\perp \frac{(\mathbf{E}\cdot \mathbf{B})}{E_\perp^2}
\left(\frac{c}{2\bar{v}}\ln\frac{c+\bar{v}}{c-\bar{v}} -1\right).
\label{new-drift-topol}
\end{eqnarray}
This is a new topological contribution, associated with the drift of plasma. It is induced when
there are both parallel and perpendicular components of the electric field. While it is intimately
connected with the chiral magnetic effect, it is not just a Lorentz boosted form of it in the laboratory frame.

To the first order in gradients and fields, the magnetization current is given by
\begin{equation}
\mathbf{j}^{\rm (curl)}_{\lambda}= \frac{\lambda c}{24\pi^2} \bm{\nabla}  \times
\left[\frac{\bar{\mathbf{v}}}{\bar{v}^3}\left(3\mu_\lambda^2+\pi^2T^2\right)
\left(\frac{c\bar{v}}{c^2-\bar{v}^2}-\frac{1}{2} \ln\frac{c+\bar{v}}{c-\bar{v}}\right)\right] .
\label{mag-current-drift}
\end{equation}
An interesting byproduct of this result is that the drift may induce a nonzero
magnetization in a chirally asymmetric plasma.

Finally, the last (dissipative) term in the complete expression for the current
(\ref{current-lambda-1}) reads
\begin{eqnarray}
\mathbf{j}^{(3)}_{\lambda}&=& \sum_{i} {\rm sign}(e_i)
\int\frac{d^3p}{(2\pi)^3}\, \mathbf{v}  \delta f^{i, (1)}_{\lambda}
 =\frac{c^2 \tau (3\mu_{\lambda}^2+\pi^2 T^2)}{12 \pi^2 }
 \left( e\frac{\left(\mathbf{E}\cdot \mathbf{B} \right) \mathbf{B} }{B^2} -\frac{\partial \mu_{\lambda}}{\partial \mathbf{x}} \right)
g_0
\nonumber\\
 & - & \frac{c \tau(\mu_{\lambda}^3+\pi^2 T^2\mu_{\lambda})}{4\pi^2} \left[
\frac{\bar{\mathbf{v}}(\bm{\nabla}\cdot\bar{\mathbf{v}})+ (\bar{\mathbf{v}}\cdot\bm{\nabla})\bar{\mathbf{v}}
+\bar{v}\bm{\nabla}\bar{v}}{\bar{v}} g_1
+\frac{\bar{\mathbf{v}}(\bar{\mathbf{v}}\cdot\bm{\nabla}\bar{v})}{\bar{v}^2} g_2
+\frac{2}{3(c^2-\bar{v}^2)^2}\frac{\partial \bar{\mathbf{v}}}{\partial t }
+\frac{8\bar{v}\bar{\mathbf{v}}}{3(c^2-\bar{v}^2)^3} \frac{\partial \bar{v}}{\partial t }
\right],
\label{currents-flow}
\end{eqnarray}
where we used the shorthand notations for the following functions of $\bar{v}/c$:
\begin{eqnarray}
g_0 &\equiv& \frac{c^3}{2}  \int_{-1}^{1} \frac{(1-\xi^2) d\xi}{(c-\bar{v} \xi)^3}
= \frac{c^3}{\bar{v}^3} \left(\frac{c\bar{v}}{c^2-\bar{v}^2}-\frac{1}{2} \ln\frac{c+\bar{v}}{c-\bar{v}}\right),\\
g_1 &\equiv&  \frac{c^4}{2} \int_{-1}^{1}\frac{\xi(1-\xi^2)d\xi}{(c-\bar{v}\xi)^4}
=\frac{c^4}{3\bar{v}^4}\left(\frac{c\bar{v}(5\bar{v}^2-3c^2)}{(c^2-\bar{v}^2)^2}
+\frac{3}{2}\ln\frac{c+\bar{v}}{c-\bar{v}}\right) ,\\
g_2 &\equiv&  \frac{c^4}{2} \int_{-1}^{1}\frac{\xi(5\xi^2-3)d\xi}{(c-\bar{v}\xi)^4}
=\frac{c^4}{3\bar{v}^4}\left(\frac{c\bar{v}(15c^4-40c^2\bar{v}^2+33\bar{v}^4)}{(c^2-\bar{v}^2)^3}
+\frac{15}{2}\ln\frac{c+\bar{v}}{c-\bar{v}}\right).
 \end{eqnarray}
 Note that, in the limit of small drift velocities $\bar{v}/c\to0$, these
 functions are nonsingular: $g_0\simeq {2}/{3} +O\left(\bar{v}^2/c^2\right)$,
 $g_1\simeq O\left(\bar{v}/c\right)$, and
 $g_2\simeq O\left(\bar{v}^3/c^3\right)$.

The result in Eq.~(\ref{CME-current}) renders the standard chiral separation and chiral
magnetic effect currents. In addition, Eq.(\ref{Hall-drift}) gives the following currents due to the Hall
effect as well as its generalization to the case of axial current:
\begin{eqnarray}
\mathbf{j}&=&c n^{\rm{(0)}}  \frac{\mathbf{E} \times \mathbf{B}}{B^2}
=\frac{\mu (\mu^2+3\mu_5^2+\pi^2T^2)}{3\pi^2 c^2\left[1-(\bar{v}/c)^2\right]^2} \frac{\mathbf{E} \times \mathbf{B}}{B^2} ,
\label{density-electric-current-Hall}\\
\mathbf{j}_5&=&c n^{\rm{(0)}}_5  \frac{\mathbf{E} \times \mathbf{B}}{B^2}
= \frac{\mu_5 (\mu_5^2+3\mu^2+\pi^2T^2)}{3\pi^2 c^2\left[1-(\bar{v}/c)^2\right]^2}\frac{\mathbf{E} \times \mathbf{B}}{B^2} .
\label{density-axial-current-Hall}
\end{eqnarray}
These have the expected structure and are not very surprising. Much more surprising are 
the new contributions to current densities in Eq.~(\ref{new-drift-topol}). The corresponding 
currents appear to be of topological origin. Indeed, they appear due to the presence of the 
Berry connection in the definition of current~\eqref{electric-current} and do not depend on
temperature or the relaxation time $\tau$. They give rise to the following electric current and 
axial current densities:
\begin{eqnarray}
\mathbf{j}&=&\frac{e \mu_{5}}{2\pi^2 c}  \mathbf{E}_\perp \frac{(\mathbf{E}\cdot \mathbf{B})}{E_\perp^2}
\left(\frac{c}{2\bar{v}}\ln\frac{c+\bar{v}}{c-\bar{v}} -1\right)
\stackrel{{\scriptstyle E_\perp\to 0}}{\longrightarrow}
\frac{e \mu_{5}}{6\pi^2 c}  \mathbf{E}_\perp \frac{(\mathbf{E}\cdot \mathbf{B})}{B^2} ,
\label{density-electric-current-drift-topol}\\
\mathbf{j}_5&=&\frac{e \mu}{2\pi^2 c}  \mathbf{E}_\perp \frac{(\mathbf{E}\cdot \mathbf{B})}{E_\perp^2}
\left(\frac{c}{2\bar{v}}\ln\frac{c+\bar{v}}{c-\bar{v}} -1\right)
\stackrel{{\scriptstyle E_\perp\to 0}}{\longrightarrow}
\frac{e \mu}{6\pi^2 c}  \mathbf{E}_\perp \frac{(\mathbf{E}\cdot \mathbf{B})}{B^2}  .
\label{density-axial-current-drift-topol}
\end{eqnarray}
These currents resemble the chiral magnetic/separation currents, but flow perpendicularly to the
magnetic field. They are of the first order in electromagnetic fields and appear to be nondissipative. 
The fact that these currents are proportional to $\mathbf{E}\cdot\mathbf{B}$ may hint at their 
possible connection with the chiral anomaly. We also observe that currents 
(\ref{density-electric-current-drift-topol}) and (\ref{density-axial-current-drift-topol}) are deeply 
connected with the existence of the plasma drift. They exist only when both the perpendicular
and parallel components of the electric field are nonvanishing. This latter suggests that the new 
topological currents cannot be eliminated, or reduced to the chiral magnetic/separation currents, 
by a simple boost transformation.

The Ohm's current in Eq.~(\ref{currents-flow}) has only the longitudinal projection with respect to the 
magnetic field. This is due to the fact that the perpendicular component of the electric field is exactly 
accounted in the drift velocity. In addition to the standard Ohm's and diffusion currents, given by the 
first term in Eq.~(\ref{currents-flow}), there are also other dissipative contributions in 
$\mathbf{j}^{(3)}_{\lambda}$, which are associated with the inhomogeneity of the drift flow. 
These appear to be connected with a nonzero viscosity of chiral plasma.

To linear order in electromagnetic fields and derivatives, the continuity equation reads
\begin{eqnarray}
\frac{\partial n^{(0)}_\lambda}{\partial \mu_\lambda} \left( \frac{\partial \mu_\lambda}{\partial t} + \bar{\mathbf{v}} \cdot
\frac{\partial \mu_\lambda}{\partial \mathbf{x}}  \right)
+ \frac{\partial n^{(0)}_\lambda}{\partial \bar{v}} \left( \frac{\partial \bar{v}}{\partial t} + \bar{\mathbf{v}} \cdot
\frac{\partial \bar{v}}{\partial \mathbf{x}}  \right)
+n^{\rm{(0)}}_\lambda \bm{\nabla} \cdot \bar{\mathbf{v}}
=0
\end{eqnarray}
or equivalently,
\begin{equation}
\left(\frac{3(\mu^2+\mu^2_5)}{\pi^2T^2}+1\right)\left(\frac{\partial \mu}{\partial t}
+ \bar{\mathbf{v}} \cdot \frac{\partial \mu}{\partial \mathbf{x}}\right)
+\frac{6\mu\mu_5}{\pi^2T^2}\left(\frac{\partial \mu_5}{\partial t}
+ \bar{\mathbf{v}} \cdot \frac{\partial \mu_5}{\partial \mathbf{x}}\right)
+\mu\left(\frac{\mu^2+3\mu^2_5}{\pi^2T^2}+1\right)
\left[\frac{4\bar{v}}{c^2-\bar{v}^2}\left( \frac{\partial \bar{v}}{\partial t} + \bar{\mathbf{v}} \cdot
\frac{\partial \bar{v}}{\partial \mathbf{x}}  \right) +  \bm{\nabla} \cdot \bar{\mathbf{v}}\right]=0,
\end{equation}
\begin{equation}
\left(\frac{3(\mu^2+\mu^2_5)}{\pi^2T^2}+1\right)\left(\frac{\partial \mu_5}{\partial t}
+ \bar{\mathbf{v}} \cdot \frac{\partial \mu_5}{\partial \mathbf{x}}\right)
+\frac{6\mu\mu_5}{\pi^2T^2}\left(\frac{\partial \mu}{\partial t}
+ \bar{\mathbf{v}} \cdot \frac{\partial \mu}{\partial \mathbf{x}}\right)
+\mu_5\left(\frac{\mu^2_5+3\mu^2}{\pi^2T^2}+1\right)
\left[\frac{4\bar{v}}{c^2-\bar{v}^2}\left( \frac{\partial \bar{v}}{\partial t} + \bar{\mathbf{v}} \cdot
\frac{\partial \bar{v}}{\partial \mathbf{x}}  \right) +  \bm{\nabla} \cdot \bar{\mathbf{v}}\right]=0.
\end{equation}
In the complete set of the anomalous Maxwell equations, these two relations determine
how local equilibrium electric and chiral chemical potentials evolve in a self-consistent way
in a chiral plasma.

\section{Expansion in powers of electromagnetic fields and derivatives}
\label{perturbation-fields}

In the previous section, we studied the chiral plasma in the drifting state. This allowed us to account
for the plasma drift exactly in the zeroth approximation of the distribution function. In this section, we
consider the case where neutral particles exert a substantial drag on the system so that one can use
the Fermi-Dirac distribution function (\ref{equilibrium-function}) as the equilibrium distribution function.
We will treat both electric and magnetic fields as perturbations. From a physics viewpoint, this
is the regime of a large collision rate \cite{Krall,Abrikosov}. In this case, an expansion in powers of both
electric and magnetic fields is justified. Since the calculations in this case become much simpler
than those in the previous section, we will derive the expressions for the densities and currents to
the second order in electromagnetic fields and derivatives. For some earlier studies using the 
effective action formalism, see also Ref.~\cite{Megias:2014mba}.

\subsection{Distribution function to quadratic order in electromagnetic fields and derivatives}
\label{quadratic}

We seek the solution to Eq.~(\ref{approxBoltz-collision-integral}) in the form of a series in powers of electromagnetic fields, i.e.,
\begin{equation}
f _{\lambda} =f^{\rm (eq)}_{\lambda}+\delta f^{(1)}_{\lambda}+\delta f^{(2)}_{\lambda}+\cdots
\end{equation}
and treat each space/time derivative as an extra power of electromagnetic field.
By substituting the above ansatz in Eq.~(\ref{approxBoltz-collision-integral}) and keeping the terms up to
linear order in electromagnetic fields, we obtain
\begin{equation}
\frac{D_{\lambda}}{T}\frac{\partial \mu_{\lambda}}{\partial t }
-\frac{D_{\lambda}}{T}  e(\mathbf{E} \cdot  \mathbf{v} )
+ \frac{D_{\lambda}}{T} \mathbf{v}\cdot \frac{\partial \mu_{\lambda}}{\partial \mathbf{x}}
+\frac{\delta f^{(1)}_{\lambda}}{\tau} =0,
\end{equation}
where we used the notation $D_{\lambda}(\mu_{\lambda}) =e^{(cp-\mu_{\lambda})/T}/(e^{(cp-\mu_{\lambda})/T}+1)^2$.
The above equation is satisfied when
\begin{equation}
\delta f^{(1)}_{\lambda}=\frac{\tau D_{\lambda}}{T} \left( e(\mathbf{E}_{\lambda} \cdot  \mathbf{v} )
- \frac{\partial \mu_{\lambda}}{\partial t } \right) .
\end{equation}
Here, by definition, $\mathbf{E}_{\lambda}\equiv \mathbf{E}-e^{-1}\partial \mu_{\lambda}/\partial \mathbf{x}$.
By making use of this result, we derive the following formal results for the densities and currents:
\begin{eqnarray}
n_\lambda  &=& \sum_{i} {\rm sign}(e_i)
\int \frac{d^3p}{(2\pi)^3}\left(f^{i\,\rm{(eq)}}_{\lambda}+ \delta f^{i, (1)}_{\lambda}
+\frac{e_i}{c} (\mathbf{B}\cdot\mathbf{\Omega}^i_{\lambda}) f^{i\,\rm{(eq)}}_{\lambda}
\right)=\frac{\mu_{\lambda}\left(\mu_{\lambda}^2+\pi^2 T^2\right) }{6\pi^2 c^3}
- \frac{\tau \left(3\mu_{\lambda}^2+\pi^2 T^2\right)  }{6 \pi^2 c^3}  \frac{\partial \mu_{\lambda}}{\partial t } ,
\label{density-lambda-1a} \\
\mathbf{j}_{\lambda} &=&  \sum_{i} {\rm sign}(e_i)
\int\frac{d^3p}{(2\pi)^3}\,\left[ \left(\mathbf{v} +e_i\mathbf{E}\times\mathbf{\Omega}^i_{\lambda}
+\frac{e_i}{c}\mathbf{B}(\mathbf{v}\cdot\mathbf{\Omega}^i_{\lambda})\right)f^{i\,\rm{(eq)}}_{\lambda}
+ \mathbf{v}  \delta f^{i, (1)}_{\lambda}
 \right] = \frac{\lambda e\mathbf{B}\mu_{\lambda}}{4\pi^2 c}
+ \frac{\tau e \mathbf{E}_{\lambda} \left(3\mu_{\lambda}^2+\pi^2 T^2\right)  }{18\pi^2 c} ,
\label{current-lambda-1a}
\end{eqnarray}
where $ {\rm sign}(e_i)$ was included in order to correctly account for the contribution of antiparticles.
As is clear, to this order, the magnetization current in Eq.~(\ref{mag-current}) does not contribute.
Then, by making use of the continuity equation, to linear order in the fields and derivatives we obtain
\begin{eqnarray}
&&\frac{3\mu_{\lambda}^2 +\pi^2 T^2}{6\pi^2 c^3} \frac{\partial \mu_{\lambda}}{\partial t }= 0 .
\end{eqnarray}
This result implies that the time derivative of $\mu_{\lambda}$ unlike the case of the drifting state considered
in the previous section vanishes to linear order. In fact, as we will see below, it is of second order in fields.
Taking this into account, the leading order correction to the  distribution function takes the following form:
\begin{equation}
\delta f^{(1)}_{\lambda} =\frac{e c \tau D_{\lambda}}{T} (\mathbf{E}_{\lambda} \cdot  \hat{\mathbf{p}} )  ,
\qquad  \hat{\mathbf{p}} =\frac{\mathbf{p}}{p},
\label{delta-f-1}
\end{equation}
where we also took into account that $\mathbf{v} = c\hat{\mathbf{p}}$.

Before proceeding to the derivation of the quadratic correction to the distribution function, let us note the
following results:
\begin{eqnarray}
\frac{\partial \delta f^{(1)}_{\lambda}}{\partial t} &=& \frac{e c \tau D_{\lambda}}{T} \frac{\partial }{\partial t}
\left(\mathbf{E}_{\lambda} \cdot \hat{\mathbf{p}}\right) ,\\
\frac{\partial \delta f^{(1)}_{\lambda}}{\partial \mathbf{p}}
&=& -\frac{e c^2 \tau D_{\lambda} }{T^2} \left(1-2f^{\rm (eq)}_{\lambda}\right)\hat{\mathbf{p}}
\left( \mathbf{E}_{\lambda} \cdot  \hat{\mathbf{p}}\right)
+ \frac{\tau D_{\lambda}}{T} \frac{e c}{p} \left[  \mathbf{E}_{\lambda} - \hat{\mathbf{p}} (\mathbf{E}_{\lambda}\cdot \hat{\mathbf{p}} )
 \right], \\
\frac{\partial \delta f^{(1)}_{\lambda}}{\partial \mathbf{x}}
&=& \frac{e c \tau D_{\lambda}}{T^2}  \left(1-2f^{\rm (eq)}_{\lambda}\right)\left(\mathbf{E}_{\lambda} \cdot \hat{\mathbf{p}}\right)
\frac{\partial \mu_{\lambda}}{\partial \mathbf{x}}
+\frac{e c \tau D_{\lambda}}{T} \frac{\partial }{\partial \mathbf{x}}  \left( \mathbf{E}_{\lambda} \cdot  \hat{\mathbf{p}} \right) .
\end{eqnarray}
To quadratic order in electromagnetic fields, the solution $\delta f^{(2)}_{\lambda} $ to the
kinetic equation takes the form:
\begin{eqnarray}
\delta f^{(2)}_{\lambda} &=&
-\frac{\tau D_{\lambda}}{T}\frac{\partial \mu_{\lambda}}{\partial t }
+\frac{e^2 c^2 \tau^2 D_{\lambda}}{T^2}  \left(1-2f^{\rm (eq)}_{\lambda}\right)\left( \mathbf{E}_{\lambda} \cdot \hat{\mathbf{p}} \right)^2
-\frac{e c \tau^2 D_{\lambda}}{T}
\left(\frac{\partial }{\partial t}  +c \hat{\mathbf{p}} \cdot  \frac{\partial }{\partial \mathbf{x}} \right)
\left(\mathbf{E}_{\lambda} \cdot \hat{\mathbf{p}}\right)
\nonumber\\
&&
+\frac{e^2 \tau D_{\lambda}}{T} (\mathbf{E}_{\lambda}\cdot\mathbf{B})(\mathbf{\Omega}_{\lambda} \cdot \hat{\mathbf{p}})
-\frac{e^2 c \tau^2 D_{\lambda}}{pT}\hat{\mathbf{p}} \cdot\left( \mathbf{B} \times \mathbf{E}_{\lambda} \right)
- \frac{e^2 c \tau^2 D_{\lambda}}{p T} \left[  ( \mathbf{E} \cdot  \mathbf{E}_{\lambda})
 - ( \mathbf{E} \cdot  \hat{\mathbf{p}}) (\mathbf{E}_{\lambda}\cdot \hat{\mathbf{p}} ) \right]
 \nonumber\\
&& -\frac{e^2 \tau D_{\lambda}}{T} (\mathbf{B}\cdot\mathbf{\Omega}_{\lambda}) \left( \mathbf{E}_{\lambda} \cdot  \hat{\mathbf{p}} \right)
-\frac{e \tau D_{\lambda}}{T}(\mathbf{E}\times \mathbf{\Omega}_{\lambda})\cdot\frac{\partial \mu_{\lambda}}{\partial \mathbf{x}}.
\end{eqnarray}
Having determined the corrections to the local equilibrium distribution function in the first and second order in electromagnetic
field and derivatives, it is not difficult to find the corresponding charge and current densities.

\subsection{Equations for the chemical potentials}
\label{equation-potentials}

We have the following results for the densities and currents:
\begin{eqnarray}
n_\lambda  &=& \sum_{i} {\rm sign}(e_i)
\int \frac{d^3p}{(2\pi)^3}\left(f^{i\,\rm{(eq)}}_{\lambda}+ \delta f^{i, (1)}_{\lambda}
+\frac{e_i}{c} (\mathbf{B}\cdot\mathbf{\Omega}^i_{\lambda}) f^{i\,\rm{(eq)}}_{\lambda}
+\frac{e_i}{c} (\mathbf{B}\cdot\mathbf{\Omega}^i_{\lambda}) \delta f^{i, (1)}_{\lambda}
+\delta f^{i, (2)}_{\lambda}\right)\nonumber\\
&=& \frac{\mu_{\lambda}\left(\mu_{\lambda}^2+\pi^2 T^2\right) }{6\pi^2 c^3}
- \frac{\tau \left(3\mu_{\lambda}^2+\pi^2 T^2\right)  }{6 \pi^2 c^3}  \frac{\partial \mu_{\lambda}}{\partial t }
-\frac{e \tau^2 \left(3\mu_{\lambda}^2+\pi^2 T^2\right)  }{18\pi^2 c}  \nabla \cdot\mathbf{E}_{\lambda}
+\frac{\lambda e^2 \tau (\mathbf{E}_{\lambda}\cdot\mathbf{B})}{4\pi^2 c}
-\frac{e\tau^2 \mu_{\lambda}}{3\pi^2 c}\left( \mathbf{E}_{\lambda} \cdot \frac{\partial \mu_{\lambda}}{\partial \mathbf{x}}\right),
\nonumber \\
&&
\label{density-lambda-2} \\
\mathbf{j}_{\lambda} &=&  \sum_{i} {\rm sign}(e_i)
\int\frac{d^3p}{(2\pi)^3}\,\left[ \left(\mathbf{v} +e_i\mathbf{E}\times\mathbf{\Omega}^i_{\lambda}
+\frac{e_i}{c}\mathbf{B}(\mathbf{v}\cdot\mathbf{\Omega}^i_{\lambda})\right)f^{i\,\rm{(eq)}}_{\lambda}
+\left(\mathbf{v} +e_i\mathbf{E}\times\mathbf{\Omega}^i_{\lambda}  \right)\delta f^{i, (1)}_{\lambda}
+ \mathbf{v}  \delta f^{i, (2)}_{\lambda} \right]+\mathbf{j}^{\rm (curl)}_{\lambda}\nonumber\\
&=&\frac{\lambda e\mathbf{B}\mu_{\lambda}}{4\pi^2 c}
+ \frac{\tau e \mathbf{E}_{\lambda} \left(3\mu_{\lambda}^2+\pi^2 T^2\right)  }{18\pi^2 c}
-\frac{e \tau^2 \left(3\mu_{\lambda}^2+\pi^2 T^2\right)}{18 \pi^2 c}\frac{\partial \mathbf{E} }{\partial t}
-\frac{e^2\tau^2 \mu_{\lambda}}{6\pi^2} (\mathbf{B}\times \mathbf{E}_{\lambda})+ \frac{\lambda e \tau  }{12\pi^2}
\bm{\nabla}  \times \left(\mu_{\lambda} \mathbf{E} \right).
\label{current-lambda-2}
\end{eqnarray}  
To this quadratic order, we had to also take into account the magnetization current in
Eq.~(\ref{mag-current}). The corresponding additional contribution is the last term in Eq.~(\ref{current-lambda-2}).
It can be rewritten in an equivalent form as follows:
\begin{equation}
\mathbf{j}_{\lambda}^{\rm (curl)}
= \frac{\lambda e \tau  }{12\pi^2} \mu_{\lambda} \bm{\nabla}  \times \mathbf{E}
- \frac{\lambda e \tau  }{12\pi^2}  \mathbf{E}  \times \frac{\partial \mu_{\lambda}}{\partial \mathbf{x}}=
-\frac{\lambda e \tau\mu_{\lambda}}{12\pi^2c}\frac{\partial \mathbf{B}}{\partial t}
- \frac{\lambda e \tau  }{12\pi^2}  \mathbf{E}  \times \frac{\partial \mu_{\lambda}}{\partial \mathbf{x}},
\end{equation}
where we used the Maxwell equation (\ref{vector-equation}) in the last equality.
In the continuity equation, of course, such a current plays no role. By substituting the results in
Eqs.~(\ref{density-lambda-2}) and (\ref{current-lambda-2}) into the continuity equation, we
obtain the sought equations for the chemical potentials
\begin{equation}
\frac{3\mu_{\lambda}^2 +\pi^2 T^2}{6\pi^2 c^3} \frac{\partial \mu_{\lambda}}{\partial t }
+ \frac{\tau e \left(3\mu_{\lambda}^2+\pi^2 T^2\right)  }{18\pi^2 c} \nabla \cdot\mathbf{E}_{\lambda}
+ \frac{\tau e  \mu_{\lambda} }{3\pi^2 c} \left(\mathbf{E}_{\lambda}\cdot \frac{\partial \mu_{\lambda}}{\partial \mathbf{x}}\right)
=\frac{\lambda  e^2\mathbf{E}_{\lambda}\cdot\mathbf{B} }{4\pi^2c} .
\end{equation}
These are equivalent to the following equations for the electric and axial charge chemical potentials:
\begin{eqnarray}
&&(3\mu^2+3\mu_5^2 +\pi^2 T^2)\left[\frac{\partial \mu}{\partial t }
- \frac{\tau c^2 }{3} \nabla^2 \mu + \frac{\tau e c^2 }{3} \nabla \cdot\mathbf{E} \right]
+6\mu \mu_5 \left[\frac{\partial \mu_5}{\partial t } - \frac{\tau c^2 }{3} \nabla^2 \mu_5\right]
+ \frac{3}{2} e c^2  \left(\mathbf{B}\cdot \frac{\partial \mu_5}{\partial \mathbf{x}}\right)
\nonumber\\
&&
+  2 \tau \mu  c^2  \left[e\mathbf{E} \cdot \frac{\partial \mu}{\partial \mathbf{x}}
- \left(\frac{\partial \mu }{\partial \mathbf{x}}\right)^2
- \left(\frac{\partial \mu_5}{\partial \mathbf{x}}\right)^2\right]
+  2 \tau \mu_5 c^2  \left[e\mathbf{E} \cdot \frac{\partial \mu_5}{\partial \mathbf{x}}
-2\frac{\partial \mu }{\partial \mathbf{x}}\cdot \frac{\partial \mu_5}{\partial \mathbf{x}}\right]
=0
,\\
&&(3\mu^2+3\mu_5^2 +\pi^2 T^2)\left[\frac{\partial \mu_5}{\partial t }
- \frac{\tau c^2 }{3} \nabla^2 \mu_5 \right]
+6\mu \mu_5 \left[\frac{\partial \mu}{\partial t } - \frac{\tau c^2 }{3} \nabla^2 \mu + \frac{\tau e c^2 }{3} \nabla \cdot\mathbf{E} \right]
+ \frac{3}{2} e c^2  \left(\mathbf{B}\cdot \frac{\partial \mu}{\partial \mathbf{x}}\right)
\nonumber\\
&&
+  2 \tau \mu_5  c^2  \left[e\mathbf{E} \cdot \frac{\partial \mu}{\partial \mathbf{x}}
- \left(\frac{\partial \mu }{\partial \mathbf{x}}\right)^2
- \left(\frac{\partial \mu_5}{\partial \mathbf{x}}\right)^2\right]
+  2 \tau \mu c^2  \left[e\mathbf{E} \cdot \frac{\partial \mu_5}{\partial \mathbf{x}}
-2\frac{\partial \mu }{\partial \mathbf{x}}\cdot \frac{\partial \mu_5}{\partial \mathbf{x}}\right] =0 .
\end{eqnarray}
These equations determine how local equilibrium electric and chiral chemical potentials evolve
in an inhomogeneous chirally asymmetric plasma. In order to complete our derivation of
anomalous Maxwell equations for inhomogeneous chiral plasma, we should find the explicit
expressions for the currents to the second order in electromagnetic field and derivatives.

To this order, the electric current and axial current densities are 
\begin{equation}
\mathbf{j}=\frac{e\mathbf{B}\mu_5}{2\pi^2 c}
+ \sum_{\lambda}\frac{\tau e \mathbf{E}_{\lambda} \left(3\mu_{\lambda}^2+\pi^2 T^2\right)  }{18\pi^2 c}
-\sum_{\lambda}\frac{e \tau^2 \left(3\mu_{\lambda}^2+\pi^2 T^2\right)}{18 \pi^2 c}\frac{\partial \mathbf{E} }{\partial t}
-\sum_{\lambda}\frac{e^2\tau^2 \mu_{\lambda}}{6\pi^2} (\mathbf{B}\times \mathbf{E}_{\lambda})
+ \sum_{\lambda}\frac{\lambda e \tau  }{12\pi^2} \bm{\nabla}  \times \left(\mu_{\lambda} \mathbf{E} \right),
\label{density-electric-current-quadratic}
\end{equation}
\begin{equation}
\mathbf{j}_5=\frac{e\mathbf{B}\mu}{2\pi^2 c}
+ \sum_{\lambda}\lambda \frac{\tau e \mathbf{E}_{\lambda} \left(3\mu_{\lambda}^2+\pi^2 T^2\right)  }{18\pi^2 c}
-\sum_{\lambda}\lambda \frac{e \tau^2 \left(3\mu_{\lambda}^2+\pi^2 T^2\right)}{18 \pi^2 c}\frac{\partial \mathbf{E} }{\partial t}
-\sum_{\lambda}\lambda  \frac{e^2\tau^2 \mu_{\lambda}}{6\pi^2} (\mathbf{B}\times \mathbf{E}_{\lambda})
+ \sum_{\lambda}\frac{ e \tau  }{12\pi^2} \bm{\nabla}  \times \left(\mu_{\lambda} \mathbf{E} \right),
\label{density-axial-current-quadratic}
\end{equation}
while the electric and axial charge densities are
\begin{eqnarray}
n&=&\sum_{\lambda} \frac{\mu_{\lambda}\left(\mu_{\lambda}^2+\pi^2 T^2\right) }{6\pi^2 c^3}
- \sum_{\lambda}\frac{\tau \left(3\mu_{\lambda}^2+\pi^2 T^2\right)  }{6 \pi^2 c^3}  \frac{\partial \mu_{\lambda}}{\partial t }
-\sum_{\lambda}\frac{e \tau^2 \left(3\mu_{\lambda}^2+\pi^2 T^2\right)  }{18\pi^2 c}  \nabla \cdot\mathbf{E}_{\lambda}
\nonumber\\
&&
-\frac{e \tau}{2\pi^2 c} \left(\mathbf{B}\cdot \frac{\partial \mu_5}{\partial \mathbf{x}} \right)
-\sum_{\lambda}\frac{e\tau^2 \mu_{\lambda}}{3\pi^2 c}\left( \mathbf{E}_{\lambda} \cdot \frac{\partial \mu_{\lambda}}{\partial \mathbf{x}}\right)
,
\label{density-electric-quadratic}\\
n_5&=&\sum_{\lambda}\lambda \frac{\mu_{\lambda}\left(\mu_{\lambda}^2+\pi^2 T^2\right) }{6\pi^2 c^3}
- \sum_{\lambda}\lambda \frac{\tau \left(3\mu_{\lambda}^2+\pi^2 T^2\right)  }{6 \pi^2 c^3}  \frac{\partial \mu_{\lambda}}{\partial t }
-\sum_{\lambda}\lambda \frac{e \tau^2 \left(3\mu_{\lambda}^2+\pi^2 T^2\right)  }{18\pi^2 c}  \nabla \cdot\mathbf{E}_{\lambda}
\nonumber\\
&&+\frac{e \tau }{2\pi^2 c}\left( e\mathbf{E}-  \frac{\partial \mu}{\partial \mathbf{x}}\right) \cdot\mathbf{B}
-\sum_{\lambda}\lambda \frac{e\tau^2 \mu_{\lambda}}{3\pi^2 c}\left( \mathbf{E}_{\lambda} \cdot
\frac{\partial \mu_{\lambda}}{\partial \mathbf{x}}\right).
\label{density-axial-quadratic}
\end{eqnarray}
Since electric and chiral chemical potentials are much smaller than temperature in a primordial plasma, it is useful to determine
the charge and current densities in the high-temperature limit.

\subsection{Electric and chiral currents in the high-temperature limit}
\label{currents}

In the high-temperature limit, the electric current and axial current densities are
\begin{eqnarray}
\mathbf{j}& \simeq &\frac{e\mathbf{B}\mu_5}{2\pi^2 c}
+ \frac{\tau T^2   }{9 c} \left(e \mathbf{E} -\frac{\partial \mu}{\partial \mathbf{x}}\right)
- \frac{e\tau^2 T^2 }{9 c}\frac{\partial \mathbf{E}}{\partial t},
\label{density-electric-current-hight-T}\\
\mathbf{j}_5&\simeq&\frac{e\mathbf{B}\mu }{2\pi^2 c}
-\frac{\tau T^2   }{9 c} \frac{\partial \mu_5 }{\partial \mathbf{x}},
\label{density-axial-current-hight-T}
\end{eqnarray}
while the electric and axial charge densities are
\begin{eqnarray}
n&\simeq& \frac{T^2 \mu  }{3 c^3}
- \frac{\tau  T^2 }{3 c^3}  \frac{\partial \mu }{\partial t }
+\frac{\tau^2  T^2 }{9  c} \left( \nabla^2 \mu -e\nabla \cdot\mathbf{E} \right)
-\frac{e \tau}{2\pi^2 c} \left(\mathbf{B}\cdot \frac{\partial \mu_5}{\partial \mathbf{x}} \right),
\label{density-electric-hight-T}\\
n_5&\simeq& \frac{T^2 \mu_5}{3 c^3}
- \frac{\tau  T^2 }{3 c^3}  \frac{\partial \mu_5 }{\partial t }
+\frac{\tau^2  T^2 }{9 c} \nabla^2 \mu_5
+\frac{e \tau }{2\pi^2 c}\left( e\mathbf{E}-  \frac{\partial \mu}{\partial \mathbf{x}}\right) \cdot\mathbf{B}.
\label{density-axial-hight-T}
\end{eqnarray}
In the high-temperature limit, $|\mu_\lambda|\ll T$, the equations for the chiral $\mu_5$
and fermion number $\mu$ chemical potentials read
\begin{eqnarray}
&&\frac{\partial \mu}{\partial t } + \frac{3c^2}{2\pi^2T^2}e\mathbf{B}\cdot\frac{\partial \mu_5}{\partial \mathbf{x}}
-\frac{\tau c^2}{3}\,\left(\nabla^2\mu-e\,\bm{\nabla}\cdot\mathbf{E}\right)
=  0,
\label{chemical-potentials-second-ordinary}
\\
&&\frac{\partial \mu_5}{\partial t } + \frac{3c^2}{2\pi^2T^2}e\mathbf{B}\cdot\frac{\partial \mu}{\partial \mathbf{x}}
-\frac{\tau c^2}{3}\,\nabla^2\mu_5
= \frac{3e^2c^2 \mathbf{E}\cdot\mathbf{B}}{2\pi^2T^2}.
\label{chemical-potentials-second-chiral}
\end{eqnarray}
It is interesting to point that the equations of motion for the chemical potentials contain the diffusion terms,
proportional to $\nabla^2\mu_5/3$ and $\nabla^2\mu/3$, with the diffusion constant given by $\tau c^2/3$.
This is in agreement with the general arguments in Ref.~\cite{Kharzeev:2010gd}. Our derivation in the present paper
not only establishes such a term, but also leads to a formal expression for the diffusion constant in terms
of the relaxation time. In general, in the presence of a nonzero magnetic field, one expects that there are
two different diffusion terms, a longitudinal one proportional to $\partial^2_z\mu_\lambda$ and a transverse
one proportional to $\Delta_T\mu_\lambda$, see, for example, Eq.~(9) in Ref.~\cite{Burnier:2012ae}. By making
use of the chiral kinetic theory, both diffusion terms can be rigorously derived. As we see from our analysis
above, to the quadratic order in the fields, the longitudinal and transverse diffusion constants are the same.
It can be shown, however, that the longitudinal diffusion constant will have a nonzero correction of order $B^2$ that
comes from the term of the type $(\mathbf{B}\cdot{\bm \nabla})(\mathbf{B} \cdot \partial \mu_{\lambda}/\partial \mathbf{x})$.

In the case of a constant magnetic field and vanishing electric field, the above set of the equations
has a solution in the form of a (diffusive) chiral magnetic wave. Indeed, by setting $\mathbf{E}=0$
and assuming that the magnetic field $\mathbf{B}$ is constant, we find that there is a
solution that describes a diffusive chiral magnetic wave with the following dispersion relation:
\begin{equation}
\omega_{\rm CMW} = \pm \frac{3c^2}{2\pi^2T^2} \left(e \mathbf{B}\cdot\mathbf{k}\right)  -\frac{i}{3}\tau c^2 |\mathbf{k}|^2.
\label{CMW-dispersion}
\end{equation}
Note that the speed of the chiral magnetic wave is given by
\begin{equation}
v_{\rm CMW} = \frac{3c^2 |eB|}{2\pi^2T^2} \cos\theta_{\mathbf{Bk}},
\label{v-CME}
\end{equation}
where $\theta_{\mathbf{Bk}}$ is the angle between the wave vector $\mathbf{k}$ and the magnetic field.

\subsection{Explicit expressions for currents}

It is instructive to discuss the physical meaning of separate contributions in the expressions for the
electric current (\ref{density-electric-current-quadratic}) and axial current (\ref{density-axial-current-quadratic})
densities. Let us start from the  electric current density. After performing the sum over $\lambda$, we derive
\begin{eqnarray}
\mathbf{j} &=& \frac{e\mathbf{B}\mu_5}{2\pi^2c}
+\frac{\tau T^2}{9 c} \left(1+\frac{3(\mu^2+\mu_5^2)}{\pi^2 T^2}\right)
\left(e\mathbf{E}-\frac{\partial \mu}{\partial \mathbf{x}}\right)
+\frac{e^2\tau^2\mu }{3\pi^2}\mathbf{E}\times\mathbf{B}
- \frac{2e\tau  \mu \mu_5}{3\pi^2 c}\frac{\partial \mu_5}{\partial \mathbf{x}}
\nonumber\\
&&
+\frac{e\tau^2\mu}{3\pi^2}\left(\mathbf{B} \times \frac{\partial \mu}{\partial \mathbf{x}}\right)
+\frac{e\tau^2\mu_5}{3\pi^2}\left(\mathbf{B} \times \frac{\partial \mu_5}{\partial \mathbf{x}}\right)
-\frac{e \tau^2 T^2}{9 c}\left(1+\frac{3(\mu^2+\mu_5^2)}{\pi^2 T^2}\right)\frac{\partial \mathbf{E} }{\partial t}
- \frac{ e \tau \mu_{5} }{6\pi^2c} \frac{\partial \mathbf{B}}{\partial t}
- \frac{ e \tau  }{6\pi^2}  \mathbf{E}  \times \frac{\partial \mu_{5}}{\partial \mathbf{x}} .
\label{electric-current-final}
\end{eqnarray}
Obviously, the first term describes the current
of the chiral magnetic effect, which is the current induced by a nonzero chiral chemical potential along the
direction of the magnetic field. The second term combines the Ohm's and diffusion currents.
Let us note that the conductivity equals
\begin{equation}
\sigma = \frac{e^2 c^2 \tau}{3} \chi^{(0)},
\label{tau-sigma}
\end{equation}
where we introduced the shorthand notations
\begin{eqnarray}
\chi^{(0)}&=&\frac{\partial n^{(0)}}{\partial \mu} =  \frac{3\mu^2+3\mu_5^2+\pi^2T^2}{3\pi^2 c^3} ,\\
n^{(0)}&=&\sum_{\lambda} \frac{\mu_{\lambda}\left(\mu_{\lambda}^2+\pi^2 T^2\right) }{6\pi^2 c^3}
= \frac{\mu (\mu^2+3\mu_5^2+\pi^2T^2)}{3\pi^2 c^3}.
\end{eqnarray}
In order to apply our results for a deconfined quark-gluon plasma, one could fix the relaxation-time 
parameter by using the lattice results \cite{sigma-lattice-1,sigma-lattice-2} for the quark-gluon
plasma conductivity (obtained at $T=1.45 T_c$) and Eq.~(\ref{tau-sigma}):
\begin{equation}
\tau \simeq 0.37 \frac{9}{\alpha T} \simeq  375~\mbox{fm}/c \left(\frac{240~\mbox{MeV}}{T}\right).
\end{equation}
The last five terms in Eq.~(\ref{electric-current-final}) are new types of terms that are produced by
time-dependent electric and magnetic fields
and gradients of the chemical potentials and, to the best of our knowledge, have not been
discussed in the literature before. We will discuss the physical meaning of each
of them, as well as their possible implications in the subsection below.

A few words are in order about the third term in Eq.~(\ref{electric-current-final}), which is nothing else but the
celebrated Hall current. At the first sight it appears to be strange that the corresponding current is proportional
to the square of the relaxation time. Yet, we emphasize that this is a standard result in the limit of large collision
rate (small $\tau$), see for example, Sec.~6.10 in Ref.~\cite{Krall}. Moreover, the usual experimental setup for
measuring the Hall effect, in which one enforces $j_y=0$, will lead to the well-known relation between the electric
current in the $x$ direction and the electric field in the $y$ direction, i.e., $j_x \propto n^2 E_y/(\mu B)$ up to
small corrections suppressed by the second power of the magnetic field and the relaxation time \cite{Abrikosov}.
Now, the leading order term in such a result is indeed standard and independent of the relaxation time.

Similarly, after performing the sum over $\lambda$ in the expression for the axial current density in
Eq.~(\ref{density-axial-current-quadratic}), we derive
\begin{eqnarray}
\mathbf{j}_5&=&\frac{e\mathbf{B}\mu}{2\pi^2c}
-\frac{e\tau T^2 }{9  c}
 \left(1+\frac{3(\mu^2+\mu_5^2)}{\pi^2 T^2}\right)
\frac{\partial \mu_5}{\partial \mathbf{x}}
+\frac{2e^2\tau \mu\mu_5}{3\pi^2 c}\mathbf{E}
-\frac{e\tau^2\mu_5 }{3\pi^2}\mathbf{B} \times\left(e\mathbf{E}-\frac{\partial \mu}{\partial \mathbf{x}}\right)
\nonumber\\
&&
+\frac{e \tau^2 \mu  }{3\pi^2}\left( \mathbf{B} \times \frac{\partial \mu_5}{\partial \mathbf{x}} \right)
-\frac{2e \tau \mu\mu_5}{3\pi^2 c}\frac{\partial \mu}{\partial \mathbf{x}}
- \frac{2e \tau^2 \mu \mu_5}{3\pi^2 c} \frac{\partial \mathbf{E} }{\partial t}
- \frac{ e \tau \mu }{6\pi^2c} \frac{\partial \mathbf{B} }{\partial t}
- \frac{ e \tau  }{6\pi^2}  \mathbf{E}  \times \frac{\partial \mu}{\partial \mathbf{x}} .
\label{chiral-current-final}
\end{eqnarray}
The first term in $\mathbf{j}_5$ is the celebrated chiral separation effect current. The second term is
a diffusion current. The third term is the axial current associated with the chiral electric separation
effect \cite{Huang:2013iia,Pu:2014cwa}. The rest are new terms.

\subsection{New contributions to the electric current}

Let us discuss the new terms in the electric current  (\ref{electric-current-final}) connected
with inhomogeneities of the electric and axial charge densities in chiral plasma. The first of the three new types of currents is associated in
a simple way with a chiral diffusion,
\begin{equation}
\mathbf{j}_{\partial\chi} = -\frac{2e\tau \mu\mu_5 }{3\pi^2 c}\frac{\partial \mu_5}{\partial \mathbf{x}} .
\label{chi-current}
\end{equation}
It is induced in a plasma in which both the fermion number and chiral chemical potentials are nonzero.
The direction of the current coincides with the gradient $\partial \mu_5/\partial \mathbf{x}$.
The current of the second type goes perpendicularly to the magnetic field, as well as to the gradient of 
the electric/chiral chemical potential, i.e.,
\begin{equation}
\mathbf{j}_{B\times\partial} = \frac{e\tau^2\mu}{3\pi^2}\left(\mathbf{B} \times \frac{\partial \mu}{\partial \mathbf{x}}\right)
+\frac{e\tau^2\mu_5}{3\pi^2}\left(\mathbf{B} \times \frac{\partial \mu_5}{\partial \mathbf{x}}\right),
\label{B-partial-chi-current}
\end{equation}
therefore, we will call this current the \emph{Hall diffusion}. The currents in Eqs.~(\ref{chi-current}) and
(\ref{B-partial-chi-current}) occur already in absence of electric fields. The current of the third type
is driven by a time-dependent electric field,
\begin{equation}
\mathbf{j}_{\partial_t E} = -\frac{e \tau^2 T^2}{9 c}\left(1+\frac{3(\mu^2+\mu_5^2)}{\pi^2 T^2}\right)\frac{\partial \mathbf{E} }{\partial t}.
\label{partial_t_E}
\end{equation}
This current is clearly a time-dependent electric field analogue of the Ohm's current, cf. the second term in
Eq.~(\ref{electric-current-final}). Finally, we also get the following contributions due to the magnetization
current:
\begin{equation}
\mathbf{j}^{\rm (curl)} = -\frac{ e \tau \mu_{5} }{6\pi^2c} \frac{\partial \mathbf{B} }{\partial t}
- \frac{ e \tau  }{6\pi^2}  \mathbf{E}  \times \frac{\partial \mu_{5}}{\partial \mathbf{x}} .
\label{j-new-mag}
\end{equation}
The second term is very interesting. It describes a current perpendicular to the electric field and
the gradient of the axial chemical potential. Such a current resembles the anomalous Hall effect
current \cite{Sinova}, which happens in the absence of magnetic field. In the case of the chiral
plasma at hand, it describes the \emph{anomalous chiral Hall effect}.

\subsection{New contributions to the axial current}

Let us now turn to the new terms in the axial current (\ref{chiral-current-final}). The first of them, i.e.,
\begin{equation}
\mathbf{j}_{5, EB} =\frac{e^2\tau^2\mu_5 }{3\pi^2}\,\mathbf{E}\times\mathbf{B},
\label{EB-current5}
\end{equation}
is a chiral analogue of the Hall effect with the axial current induced in a medium with $\mu_5\neq 0$
\cite{Pu:2014fva}. The existence of this term is very interesting. In principle, it allows one to determine 
experimentally the sign of the chiral charge of dominant carriers in a chiral plasma. The corresponding 
sign could be extracted from the direction of $\mathbf{j}_{5}$ in orthogonal electric and magnetic fields.

The current of the second type is driven by gradients of the electric and chiral chemical potentials and a perpendicular
magnetic field, i.e.,
\begin{equation}
\mathbf{j}_{5,B\times\partial} =\frac{e \tau^2 \mu_5  }{3\pi^2}\left( \mathbf{B} \times \frac{\partial \mu}{\partial \mathbf{x}} \right)
+\frac{e \tau^2 \mu  }{3\pi^2}\left( \mathbf{B} \times \frac{\partial \mu_5}{\partial \mathbf{x}} \right).
\label{B-partial-chi-current5}
\end{equation}
Obviously, this current is a chiral analogue to the Hall diffusion current in Eq.~(\ref{B-partial-chi-current}). The last contribution to
the axial current is given by two terms
\begin{equation}
\mathbf{j}_{5,\partial_t E}=-\frac{2e \tau \mu\mu_5}{3\pi^2 c}\frac{\partial \mu}{\partial \mathbf{x}}
-\frac{2e \tau^2 \mu \mu_5}{3\pi^2 c} \frac{\partial \mathbf{E} }{\partial t}.
\label{chi-E-current5}
\end{equation}
Since current (\ref{chi-E-current5}) vanishes in a plasma where $\mu$ or $\mu_5$ equals zero, this current
defines diffusion and time-dependent electric field analogues of the current of the chiral electric separation effect given by
the third term in Eq.~(\ref{chiral-current-final}) (note that the corresponding numerical coefficients of the currents match too).
At last, the magnetization current gives the following contribution to the axial current density:
\begin{equation}
\mathbf{j}^{\rm (curl)}_5 = -\frac{ e \tau \mu }{6\pi^2c} \frac{\partial \mathbf{B} }{\partial t}
- \frac{ e \tau  }{6\pi^2}  \mathbf{E}  \times \frac{\partial \mu }{\partial \mathbf{x}} .
\label{j5-new-mag}
\end{equation}
The last term is analogous to the anomalous chiral Hall effect current, given by the last term in Eq.~(\ref{j-new-mag}).
However, the corresponding contribution to the axial current is perhaps even more interesting. It implies that,
in an electric field, the perpendicular component of the gradient of the chemical potential should lead to a
nonzero $\mathbf{j}_{5}$.

Before concluding this section, let us emphasize that the main results of this section are the explicit
expressions for the currents in an inhomogeneous chiral plasma. These expressions provide a critical ingredient in
the analysis of the anomalous Maxwell equations (\ref{Poisson-equation}) through (\ref{Maxwell-equations}),
together with the equations for time-dependent and spatially inhomogeneous chemical potentials.
The corresponding complete set of equations is a starting point for the future studies of the inverse
cascade scenarios with a realistic treatment of plasma inhomogeneities.

\section{Conclusion}
\label{Conclusion}

In this paper, by making use of the chiral kinetic equation, we derived a closed set of anomalous 
Maxwell equations relevant for the study of relativistic plasmas with chiral asymmetry and 
inhomogeneities. By utilizing an expansion in powers of electromagnetic fields and derivatives, 
we derived electric and axial currents as well as a closed set of the coordinate-space
equations for the electric (or fermion number) and chiral chemical potentials.
We studied the two regimes in which the zero order distribution
function is given by the standard Fermi-Dirac distribution function and a
boosted one. The latter realizes the drifting state where the plasma drifts
as a whole with the drift velocity perpendicular to both electric and magnetic
fields. In this case, the expansion proceeds only in powers of the component
of electric field parallel to the magnetic field whereas the perpendicular component
of electric field is taken into consideration exactly. In addition to the Hall current
for the electric current, we found its analogue for the axial current generated by the
axial density. The chiral magnetic effect current is reproduced exactly in the drifting
state. What is surprising is that we also found two additional electric and axial currents of
a possible topological origin. They resemble the chiral magnetic/separation currents,
but flow perpendicular to the magnetic field and are driven by the perpendicular component of
electric field, as well as the scalar product of the electric and magnetic fields.

While a relativistic QED electron-positron plasma provides perhaps one the best examples
of a plasma in the drifting state, quark-gluon plasma at sufficiently large temperatures may
give an example of plasma where the drift is not fully developed because electromagnetically
neutral gluons provide an essential drag on the charged particles. For such a case, we
treated both electric and magnetic fields as perturbations and derived the expressions for
the densities and currents to the second order in electromagnetic fields and derivatives.
In the special case of vanishing electric field, we found a solution in the form of a diffusive chiral
magnetic wave with the propagation speed $v_{\rm CMW} \propto |eB|/T^2$, see Eq.~(\ref{v-CME}).
The diffusion constant is given by $c^2\tau/3$, where $\tau\sim 1/\left[e^4T\ln(1/|e|)\right]$
is the relaxation time \cite{Arnold}.

In the framework used, we also derived the explicit expressions for the fermion-number and chiral densities,
as well as the corresponding currents. The results are in agreement with the continuity equations,
supplemented by the appropriate quantum anomaly term. The final results reproduce several previously
known effects. In the case of the electric current, we reproduced the Ohm's and diffusion currents,
as well as the the chiral magnetic effect. In addition, we found several new types of contributions
connected with inhomogeneities in relativistic plasma that have not been reported before. They are
the chiral diffusion and two diffusion terms of the Hall type perpendicular to the magnetic field. There
is also a term driven by a time-dependent electric field. One of the very interesting new contributions
is the anomalous chiral Hall effect current, which describes an electric current perpendicular to the
applied electric field and the gradient $\partial \mu_5/\partial \mathbf{x}$. The origin of this current
is related to the Berry curvature. We also find that there is its counterpart in the axial current density,
which is driven by the electric field and the gradient of the chemical potential. It is even more amazing
because it can be induced even in a chirally symmetric plasma.

In the case of the chiral current, we reproduced the well-known results for the chiral separation and chiral
electric separation effects. We also found an expected diffusion current and several new
types of currents including a chiral analogue of the Hall effect, which may allow one to experimentally
determine the sign of the chiral charge of dominant carriers in a chiral plasma. In addition, there are
two new types of chiral diffusion currents of the Hall type and two diffusion and time-dependent electric field analogues of the
chiral electric separation effect.

The theoretical framework of our study here is a starting point in the study of the  inverse cascade
scenario proposed in a number of recent papers. In the simplified analysis of the corresponding
dynamics, it is often assumed that a space-average currents and density correctly capture the
underlying dynamics of the inverse cascade. It is quite natural to suggest that such an assumption
may not be justified. Indeed, the underlying mechanism relies on the clear separation of the short-
and long-range modes at the scale set by the chiral chemical potential. If the latter is not uniform,
a whole window of length scales opens, where the dynamics does not have a preferred direction
of the cascade. If the underlying processes in this region would happen to enhance the degree
of inhomogeneities, the corresponding window of length scales would widen and prematurely
quench the cascade. Indirectly, this may have been suggested by a recent study in
Ref.~\cite{Buividovich:2015jfa}, although the conclusions of that study may not be conclusive. 
In fact, there are indications that the inverse cascade in the model of Ref.~\cite{Buividovich:2015jfa} 
should be realized when sufficiently large lattices are used \cite{foot1}. 

The natural continuation of the present study is a critical reexamination of the cosmological
inverse cascade scenario, in which plasma inhomogeneities are properly accounted for. The
corresponding investigation seems possible only by making extensive use of numerical methods.
That is beyond the scope of the present paper, but will be attempted in the future study and
reported elsewhere.

\acknowledgments
The authors would like to thank P.~Buividovich, C.~Manuel, E.~Megias, Shi~Pu, A.~Sadofyev, 
and N.~Yamamoto for useful comments about the results in this paper. A.B. and O.R. would like 
to thank Jurg Fr\"ohlich for many important discussions and collaboration on related matters. The 
work was supported in part by the Swiss National Science Foundation, Grant No. SCOPE IZ7370-152581.
E.V.G. is grateful to the Program of Fundamental Research of the Physics and Astronomy 
Division of the NAS of Ukraine for the support. The work of I.A.S. was supported by the U.S. 
National Science Foundation under Grant No.~PHY-1404232. S.V. is grateful to the Swiss 
National Science Foundation (individual Grant No. IZKOZ2-154984) and to Professor Ruth Durrer 
for useful discussions, her support, and kind hospitality.

\noindent 
\emph{Noted added.} --- When finishing this paper, we became aware of a partially overlapping study
by Jiunn-Wei Chen, Takeaki Ishii, Shi Pu, and Naoki Yamamoto \cite{Chen:2016xtg}.

\appendix

\section*{Appendix: Useful table integrals and relations}
\label{table-integrals}

In this appendix, for reader's convenience, we list the key table integrals and relations used in the
main text.

By making use of the shorthand notation $D_{\lambda}(\mu_{\lambda})
\equiv f^{\rm (eq)}_{\lambda}\left(1-f^{\rm (eq)}_{\lambda}\right)
=e^{(cp-\mu_{\lambda})/T}/(e^{(cp-\mu_{\lambda})/T}+1)^2$, is it straightforward
to derive the following results of integrations over the momenta:
\begin{eqnarray}
\int\frac{d^3p}{(2\pi)^3} \frac{D_{\lambda}(\mu_{\lambda}) }{p^2}
&=& \frac{T}{2\pi^2 c} \frac{1}{1+e^{-\mu_{\lambda}/T}}  ,
\label{integral-2} \\
\int\frac{d^3p}{(2\pi)^3} \frac{D_{\lambda}(\mu_{\lambda}) }{p}
&=& \frac{T^2}{2\pi^2 c^2} \ln\left(1+e^{\mu_{\lambda}/T}\right),
\label{integral-3} \\
\int\frac{d^3p}{(2\pi)^3} p^{n-2}  f^{\rm (eq)}_{\lambda}(\mu_{\lambda})
&=& -\frac{T^{n+1} \Gamma(n+1) }{2\pi^2 c^{n+1}}  \mbox{Li}_{n+1}\left(-e^{\mu_{\lambda}/T}\right),
\qquad n\geq 0,
\label{integral-3a} \\
\int\frac{d^3p}{(2\pi)^3} p^{n-2}  D_{\lambda}(\mu_{\lambda})
&=& -\frac{T^{n+1} \Gamma(n+1) }{2\pi^2 c^{n+1}}  \mbox{Li}_{n}\left(-e^{\mu_{\lambda}/T}\right),
\qquad n\geq 0,
\label{integral-3b} \\
\int\frac{d^3p}{(2\pi)^3} D_{\lambda}(\mu_{\lambda})  \left(1-2f^{\rm (eq)}_{\lambda}\right)
&=& \frac{T^3}{\pi^2 c^3} \ln\left(1+e^{\mu_{\lambda}/T}\right).
\label{integral-4}
\end{eqnarray}
When calculating the contributions of antiparticles, one also encounters similar integrals with
$D_{\lambda}(\mu_{\lambda}) \to D_{\lambda}(-\mu_{\lambda})$. In order to simplify the final expressions
for currents and densities, then, it is often useful to take into account the following relations:
\begin{eqnarray}
\frac{1}{1+e^{x}}+\frac{1}{1+e^{-x}} = 1,\\
\ln(1+e^{x})-\ln(1+e^{-x})&= & x,\\
\mbox{Li}_{2} (-e^{x}) +\mbox{Li}_{2} (-e^{-x})  &= & -\frac{x^2}{2}-\frac{\pi^2}{6} ,\\
\mbox{Li}_{3} (-e^{x}) -\mbox{Li}_{3} (-e^{-x})  &= & -\frac{x^3}{6}-\frac{\pi^2 x}{6},\\
\mbox{Li}_{4} (-e^{x}) + \mbox{Li}_{4} (-e^{-x})  &= & -\frac{x^4}{24}-\frac{\pi^2 x^2}{12}-\frac{7\pi^4}{360} .
\end{eqnarray}


\begin{thebibliography}{99}
\bibitem{footnote} We would like to thank Naoki~Yamamoto for reminding us the 
physical meaning of this term.
\end{thebibliography}
\end{document}